\documentstyle[amssymb,11pt,osid]{article}

\begin{document}

\title{Classical analogs for Rabi-oscillations, Ramsey-fringes, and
spin-echo in Josephson junctions}
\author{Jeffrey E.~Marchese \\
{\footnotesize {\it Department of Applied Science, University of California,
Davis, California 95616, USA}}\\
[2ex] Matteo Cirillo \\
{\footnotesize {\it Dipartimento di Fisica and INFM, Universit\`{a} di Roma
"Tor Vergata", I-00173 Roma, Italy}} \\
[2ex] Niels Gr{\o }nbech-Jensen \\
{\footnotesize {\it Department of Applied Science, University of California,
Davis, California 95616, USA}} }
\maketitle

\begin{abstract}
We investigate the results of recently published experiments on the quantum
behavior of Josephson circuits in terms of the classical modelling based on
the resistively and capacitively-shunted (RCSJ) junction model. Our analysis
shows evidence for a close analogy between the nonlinear behavior of a
pulsed microwave-driven Josephson junction at low temperature and low
dissipation and the experimental observations reported for the Josephson
circuits. Specifically, we demonstrate that Rabi-oscillations,
Ramsey-fringes, and spin-echo observations are not phenomena with a unique
quantum interpretation. In fact, they are natural consequences of transients
to phase-locking in classical nonlinear dynamics and can be observed in a
purely classical model of a Josephson junction when the experimental recipe
for the application of microwaves is followed and the experimental detection
scheme followed. We therefore conclude that classical nonlinear dynamics can
contribute to the understanding of relevant experimental observations of
Josephson response to various microwave perturbations at very low
temperature and low dissipation.
\end{abstract}

\section{Introduction}

In recent years the possibility of employing Josephson junctions circuitry
as solid state elements for quantum information processing has been
investigated by several groups and authors both at theoretical and
experimental levels \cite{QCQBMS}. At temperatures low enough, operating the
electronic \ Josephson devices in the regime where the thermal energy is
smaller than the energy spacing between quantum mechanical energy levels,
interpretation of experimental results in terms of quantum response has been
reported in a number of publications \cite%
{Vion_02,Martinis_02,Vion_03,Martinis_03,Chiorescu,Claudon_04,Kutsuzawa,
Wallraff,Plourde,Koch,Simmonds}. Specific key observations that are being
used to characterize the nature of a Josephson system at these extremely low
temperatures are usually linked to the system response to a variety of
pulsed microwave applications. The reason is at least tri-fold. First, due
to the nature of the Josephson effect, it is not easy to make direct
observations of the expected quantum states in the anharmonic potential of a
zero-mean voltage state Josephson system, which is the state of interest, so
convenient probes of the system are those where transitions from
zero-voltage states to non-zero states can be induced with different
probability depending on the nature of the initial zero-voltage state of the
system. Second, the analogy between a single degree of freedom quantum
Josephson system and a more traditional quantum system allows for useful
characterization tools for Josephson applications of quantum mechanical
features from, e.g., atomic systems, to infer the quantum behavior of the
system from concepts such as Rabi oscillations \cite{Rabi}, Ramsey fringes 
\cite{Ramsey}, and spin echo. Third, the control and manipulation of the
state of the system is conveniently directed through microwave frequencies
that are commensurate with the anticipated energy level spacings in the
zero-voltage Josephson state. The understanding of the Josephson response to
pulsed microwave applications is therefore critical for the interpretation
of experimental observables.

A classical nonlinear oscillator, which is subjected to pulsed time-varying
perturbations, may exhibit several phenomena that are relevant for microwave
perturbed Josephson systems. For example, an anharmonic oscillator provides
an amplitude-dependent resonance frequency, which, in turn, relates
resonance frequency directly to system energy \cite%
{Samuelsen_73,Marchese_06_1}. Also, the transient system behavior, due to an
onset of resonant microwave application to a low-dissipation Josephson
system, can give rise to low frequency amplitude and phase modulations in
the system response \cite{Marchese_06_1,Samuelsen_88,Jensen_93,Jensen_05}.
It has been shown that these phenomena in classical nonlinear dynamics are
not unlike what one would expect from a comparable quantum system responding
to a microwave signal, which is resonant with the intrinsic energy level
spacing of the potential energy surface. In fact, the simplest possible
classical Josephson system, a single Josephson junction, has been shown to
exhibit phenomena of resonant escape from the anharmonic potential well \cite%
{Jensen_04_1,Jensen_04_2,Cirillo_06}, as well as transient slow modulations
to phase-locking closely resembling experimental observations of
Rabi-oscillations and Ramsey-fringes in Josephson circuits \cite%
{Marchese_06_1,Jensen_05,Marchese_07}. The correspondence is further
substantiated by conducting numerical simulations of the classical Josephson
model in accordance with the experimental procedures, and extracting the
resulting data as it is done experimentally; namely as a statistical
probability of escape from the Josephson potential well. Also, it has been
observed that the classical model exhibits the phenomena of interest in the
limits of low temperature and low dissipation (high-Q), consistent with
reported experiments. With the demonstration of close classical analogies to
Rabi-oscillations and Ramsey-fringes, which are themselves closely related
phenomena, we here extend the investigation of the classical model to also
include a demonstration of the classical equivalent to the spin-echo
observations \cite{Vion_02,Vion_03}, since these represent yet another
manifestation of the transient modulations (Rabi-oscillations) in the
approach to a phase-locked (excited) dynamical state. The work described
here is primarily meant to demonstrate the phenomena in question in the most
simple system possible; the single Josephson junction. However, while we
will here limit ourselves to a descriptive treatment of a simple system, the
phenomena are obvious and ubiquitous to a broad range of classical Josephson
circuits, including one or more superconducting loops containing one or more
Josephson junctions manipulated through magnetic fields imposed on the loops.

\section{The Model}

The normalized classical equation for a perturbed Josephson junction within 
\[
\ddot{\varphi}+\alpha \dot{\varphi}+\sin \varphi =\eta +\varepsilon
_{s}(t)\sin (\omega _{s}t+\theta _{s})+\varepsilon _{p}(t)+n(t)\;,
\]%
where $\varphi $ is the difference between the phases of the quantum
mechanical wave functions defining the junction, $\eta $ represents the dc
bias current, and $\varepsilon _{s}(t)$, $\omega _{s}$, and $\theta _{s}$
represent microwave amplitude, frequency, and phase, respectively. All
currents are normalized to the critical Josephson current $I_{c}$, and time
is measured in units of the inverse Josephson plasma frequency, ${\omega _{0}%
}^{-1}$, where ${\omega _{0}}^{2}=2eI_{c}/\hbar C=2\pi I_{c}/\Phi _{0}C$, $C$
being the capacitance of the junction, and $\Phi _{0}=h/2e=2.07x10^{-15}$ $Wb
$ is the flux quantum. Tunneling of quasiparticles is represented by the
dissipative term, where $\alpha =\hbar \omega _{0}/2eRI_{c}$ is given by the
shunt resistance $R$, and the accompanying thermal fluctuations are defined
by the dissipation-fluctuation relationship \cite{Parisi_88} 
\begin{eqnarray}
\left\langle n(t)\right\rangle  &=&0 \\
\left\langle n(t_{1})n(t_{2})\right\rangle  &=&2\alpha {\frac{k_{B}T}{H_{J}}}%
\delta (t_{2}-t_{1})=2\alpha \Theta \delta (t_{2}-t_{1}),
\end{eqnarray}%
$T$ being the thermodynamic temperature, $H_{J}$ is the characteristic
Josephson energy $H_{J}=I_{c}\hbar /2e$, and $\Theta =k_{B}T/H_{J}$ is
thereby defined as the normalized temperature (or normalized thermal
energy). In order to give orders of magnitudes for the typical realistic
parameters we indicate that a good quality Josephson junction having a
critical current of $I_{c}=5$ $\mu A$ \ and fabricated in the standard
Nb-trilayer technology \cite{Hasuo87} may have $R$ (usually identified with
the subgap resistance) in the range $(3-6)k\Omega $ , a capacitance $%
C\approxeq 5pF$ $,$ $\nu _{0}={\omega _{0}/2\pi }\approxeq 9GHz,\alpha $ in
the range $(10^{-3}-10^{-4}),H_{J}\approxeq 16x10^{-22}$ $J$. The values of $%
R$ an $\alpha $ in the specified interval depend on the quality of the
junctions.

A current pulse for probing the state of the system is represented by $%
\varepsilon _{p}(t)$. Figure 2 sketches the signaling details relevant for
the observations of Rabi-oscillations (Fig.~2a), Ramsey-fringes (Fig.~2b),
and spin-echo (Fig.~2c). The system is initially in its zero-voltage state
(thermalized in the potential well illustrated in Figure 1b), when a
microwave is applied in a single interval or in a sequence of pulses, which
have the same reference phase $\theta _{s}$. Due to the subsequent
application of the probe pulse $\varepsilon _{p}(t)$, the system variable $%
\varphi $ may escape the potential well, and the amplitudes of $\varepsilon
_{p}(t)$ is parameterized such that probe pulse induced escape is likely if
the system energy, relative to the potential well depth, is significant at
the time immediately prior to the onset of $\varepsilon _{p}(t)$, and
unlikely otherwise. By conducting a large number of such simulations, each
with identical parameters, except for $\theta _{s}$, which is chosen from a
uniform distribution in the interval $]0,2\pi ]$, and the thermal noise
current, a statistical signature of the system energy due to the signaling
strategy given by $\varepsilon _{s}(t)$ can be inferred from the resulting
switching probability. This information can be directly obtained in
simulations by calculating the defined normalized energy 
\[
H=\frac{1}{2}\dot{\varphi}^{2}+1-\cos \varphi -\eta \varphi \;, 
\]%
where the minimum of the potential well can be represented by the energy 
\[
H_{0}=1-\sqrt{1-\eta ^{2}}-\eta \sin ^{-1}\eta \;. 
\]%
The above statistical procedure of monitoring the probability of escape is
applied experimentally, since direct measurements of the energy content in a
zero-average voltage state is not possible. We therefore mimic the
experimental procedure in our simulations in order to make direct linkage
between modeling and experimental observations of switching probabilities.
We finally notice two essential system parameters that are related to the
choice of bias point $\eta $; namely microwave frequency $\omega _{s}$ and
temperature $\Theta $. The microwave frequency should be chosen such that
resonant states can be excited in the vicinity of $\omega _{s}^{2}\approx
\omega _{l}^{2}=\sqrt{1-\eta ^{2}}$, and the thermal energy $\Theta $ should
be significantly smaller than the potential well depth $\Theta \ll \Delta U=2%
\sqrt{1-\eta ^{2}}+2\eta \sin ^{-1}\eta -\pi \eta $, ensuring that recorded
switching events are not simply a reflection of thermal activation.

\section{Simulation Details and Results}

Following the procedure of reported experiments, we record the switching
from the zero-voltage state ($\langle\dot\varphi\rangle=0$) as a result of
applying the three different recipes sketched in Figure 2. We will, unless
otherwise noted, throughout this presentation use the following parameters: $%
\eta=0.904706$, which is equivalent to a linear resonance frequency $%
\omega_l=\sqrt[4]{1-\eta^2}=0.652714$, $\alpha=10^{-4}$, which provides a
high-Q resonator, and, when temperature is applied, $\Theta=2\times10^{-4}%
\ll\Delta U=5.57\times10^{-2}$.

\subsection{Rabi-Oscillations}

The first signaling procedure is shown in Figure 2a, where the Josephson
junction (Equation (1)) is initially resting in the potential well, when the
microwave pulse is initiated. We use a microwave frequency $%
\omega_s=\omega_l $, but we note that this is not a requirement for the
observations of this work \cite{Marchese_06_1}. Figure 3 shows the evolution
for $\Theta=0$ of the junction phase $\varphi(t)$ and the system energy $%
H(t)-H_0$ for a microwave pulse with amplitude $\epsilon_s=0.00217$. Several
important features can be extracted from this figure. First, we observe how
the junction is driven into a phase-locked state by the microwave
application. Second, we observe how the junction response is modulated by a
low frequency envelope function that causes the system energy to oscillate
around the steady-state phase-locked energy. Third, comparing two different
durations of microwave pulses, we illustrate how the microwave duration
influences the switching probability when the probe pulse is applied after
the termination of the microwave field. Figures 3a and 3b show how the probe
pulse catches the system in a low energy state, resulting in a "no-switch"
count, and Figures 3c and 3d show the probe pulse catching the system in a
high energy state, resulting in a "switch" count. Thus, by varying the
duration of the microwave pulse, we can observe the slow modulation
frequency of the transient to the phase-locked state in the zero-voltage
state of the junction by observing the switching statistics by the probe
pulse application. Such statistics is shown in Figure 4a for $%
\Theta=2\times10^{-4} $, where the switching probability as a function of
microwave duration is displayed for averages of $2500$ possible switching
events. Clearly, the probability exhibits the slow modulation seen in Figure
3, and these are the classical analogues to Rabi-oscillations \cite%
{Marchese_06_1,Jensen_05}. These oscillations have a reasonably well-defined
frequency. The amplitude decays for several reasons, including the
dissipation $\alpha$, which sets a time scale for the transients to
phase-locking \cite{Marchese_06_1}, thermal effects, which provides
decoherence to the evolution, and the random phase $\theta_s$, which
dephases the different trajectories that are averaged to give Figure 4a.
Based on Figure 4a, we can now extend the analogy to quantum mechanics by
noticing that a microwave pulse of one Rabi-period $\tau_R=2\pi/\Omega_R$,
where $\Omega_R$ is the modulation (Rabi) frequency, leaves the system
energy in roughly the same state as before the microwave pulse was applied.
We therefore denote such a pulse a $2\pi$-pulse. Similarly, we define a $\pi$%
-pulse to be one of duration $\tau_R/2$, and a $\pi/2$-pulse represents a
duration $\tau_R/4$, which is the duration it takes from the microwave onset
to elevate the system to its steady state energy.

\subsection{Ramsey-Fringes}

Following the recipe outlined in, e.g., Ref.~\cite{Vion_02,Vion_03} we can
now apply a sequence of controlled microwave pulses, such as illustrated in
Figure 2b, where two $\pi/2$ pulses are applied with a delay before the
probe pulse determines if the resulting state has enough energy to be
perturbed into escape. Figure 5 illustrates, in analogy with Figure 3 for $%
\Theta=0$, the two characteristic outcomes of temporal evolutions of the
system dynamics for two different time intervals between the $\pi/2$-pulses.
The first pulse elevates the system energy to the intermediate level, where
it is left to evolve freely after the first pulse is terminated. During the
time between the two pulses, the microwave field and the junction oscillate
at almost the same frequency, but with a small detuning of the mutual phase.
Thus, this ballistic interval can produce very different outcomes when the
second $\pi/2$-pulse is applied, depending on the time interval. If the time
interval is such that the second pulse is initiated with a mutual phase
difference to the junction similar to when the first pulse terminated, then
the second pulse will continue to pump energy into the junction, elevating
its energy to the maximum at the conclusion of the second pulse. This
scenario is shown in Figures 5c and 5d, where the probe pulse triggers an
escape event. However, if the second $\pi/2$-pulse is initiated such the the
relative phase is opposite to what it was when the first pulse was
terminated, then the second pulse will attenuate the energy of the junction,
leaving it with very low energy at the conclusion of the second pulse. This
scenario is illustrated in Figures 5a and 5b, where the probe pulse cannot
induce an escape event. The switching probability as a function of the
ballistic time interval between the two $\pi/2$-pulses is shown for $%
\Theta=2\times10^{-4}$ and microwave frequency $\omega_s=\omega_l$ in Figure
4b. The clearly visible oscillations in the switching probability are the
classical equivalent of quantum mechanical Ramsey-fringes. As was the case
for the Rabi-oscillations, the visible Ramsey-fringes have a reasonably well
defined frequency $\Omega_F$ and an attenuating amplitude, where the
attenuation is due to a combination of dissipation, thermalization, and the
averaging over trajectories of different random microwave phases $\theta_s$.
The close relationship between the classical observation shown here and
quantum mechanical Ramsey-fringes can be further underlined by evaluating
the Ramsey-fringe frequency $\Omega_F$ as a function of frequency detuning
form the harmonic resonance frequency $\omega_l$. This is shown in Figure 6,
where black markers and associated error bars represent the measured
Ramsey-fringe frequency. We clearly observe the characteristic unity-slope
("V"-shape indicated by the dashed lines) dependency on the frequency
detuning from linear resonance \cite{Plourde}. An obvious deviation from a
naive expectation is that the microwave frequency for which $\Omega_F=0$ is
not exactly the linear resonance. This is due to the anharmonicity of the
potential as well as to the dissipation in the system. The Ramsey-fringe
frequency $\Omega_F$ is drifting slightly toward higher values as the $\pi/2$%
-pulse separation is increased. This is because higher oscillation
amplitudes result in lower resonance frequencies. Thus, defining the
Ramsey-fringe frequency from the first couple of oscillations (the ones with
the best resolution) will make $\Omega_F$ susceptible to the anharmonicity
of the potential. This, in turn, will make the ballistic evolution between
the two $\pi/2$-pulses commensurate with a slightly detuned microwave
frequency such that $\Omega_F=0$ for $\omega_s<\omega_l$; which is what can
be observed in Figure 6a. Figure 6b illustrates how increasing the
dissipation parameter to $\alpha=10^{-3}$ will provide a faster
linearization of the resonance frequency of the junction, resulting in the
expected shift in the characteristic point where $\Omega_F=0$ for $%
\omega_s\approx\omega_l$.

\subsection{Spin-Echo}

We finally study the result of applying the signaling protocol shown in
Figure 2c. This protocol is similar to the one resulting in Ramsey-fringes,
but a $\pi$-pulse is inserted between the two $\pi/2$-pulses in order to
mimic a 180$^\circ$ phase shift in the modulation dynamics. In analogy with
experimental observations of spin-echo in Josephson systems \cite%
{Vion_02,Vion_03} we choose a temporal separation between the two $\pi/2$%
-pulses, and observe the switching characteristics of the junction, when the
probe pulse is applied, as a function of the temporal position of the $\pi$%
-pulse, which serves as a 180$^\circ$ phase-shift from which an "echo" can
be detected. Since the system either switches or remains in the potential
well after the application of the probe, we can again demonstrate the
typical dynamics by the two cases shown for $\Theta=0$ in Figure 7. Here we
have chosen a total normalized time interval for the $\pi/2$-pulses of 2500
time units. The first case, which is displayed in Figures 7a and 7b, shows
that the application of the $\pi$-pulse only has a marginal effect on the
evolution of the phase and its energy. Since the second $\pi/2$-pulse is
initiated out of phase with the junction, the subsequent probe pulse
perturbation cannot make the system escape the potential well. In contrast,
Figures 7c and 7d show that a slightly different timing in the application
of the $\pi$-pulse will provide a resonant increase in the system
oscillation amplitude and energy. Further, the phase-relationship between
the microwave field and the junction is now such that the second $\pi/2$%
-pulse enhances the energy of the system. This results in an escape event
when the probe pulse is applied. The probability of switching as a function
of the timing of $\pi$-pulse initiation is shown in Figure 4c for $%
\Theta=2\times10^{-4}$ and $\omega_s=0.965\omega_l$. We find very clear
signatures of the oscillations in the switching probability, which can be
interpreted as the phase-dependency of the "echo" that either leaves the
system in high or low energy states. Thus, this is the classical Josephson
analogue to quantum mechanical spin-echo. We again notice that the
oscillations provide a certain frequency, which we have displayed in Figure
6a as open markers as a function of microwave frequency detuning. It is
obvious that the frequency is closely related to the Ramsey-fringe frequency 
$\Omega_F$, except near the point $\Omega_F=0$. This deviation can be
directly associated with the drift in frequency due to anharmonicity that we
mentioned above for the Ramsey-fringes, and the exact values of the
spin-echo oscillations seem to depend strongly on which part of the curve
(such as the one shown in Figure 4c) is used to calculate the frequency when
the oscillation frequency is small.

\section{Conclusions}

Just like spectroscopic resonant multi-peak switching distributions can be
observed and explained within a classical Josephson framework \cite%
{Jensen_04_1}, so can the phenomena of Rabi-oscillations, Ramsey-fringes,
and spin-echo be observed in the purely classical Josephson model by
following the recipe of the experimental observations. We have here tried to
outline this observation in the simplest possible system; namely the single
Josephson junction. This is not a direct representation of specific
experimental configurations, which are much more complex, although the
parameter values chosen here have been inspired by several relevant
experiments. Considering the simplicity of the single junction and the
modeling approach in this presentation, the overall agreement between the
classical results and experimental observations is remarkable. Not only do
we observe the abovementioned phenomena in the simple classical model, but
we also observe the magnitudes of, and relationships between, the various
oscillations in the respective switching distributions to be quite
reasonably aligned with experimental observations and expectations. We
therefore submit that much in the interpretation of the experiments on
microwave manipulations of Josephson systems at low temperature and low
dissipation can be contributed from the well known and well documented
classical Josephson dynamics.

\section{Acknowledgment}

This work was supported in part by the UC~Davis Center for Digital Security,
AFOSR grant FA9550-04-1-0171, in part by MIUR (Italy) COFIN04.

\pagebreak 
\begin{figure}[h]
\setlength{\unitlength}{0.45 cm} 
\begin{picture}(10,15)(0,0)
        \includegraphics{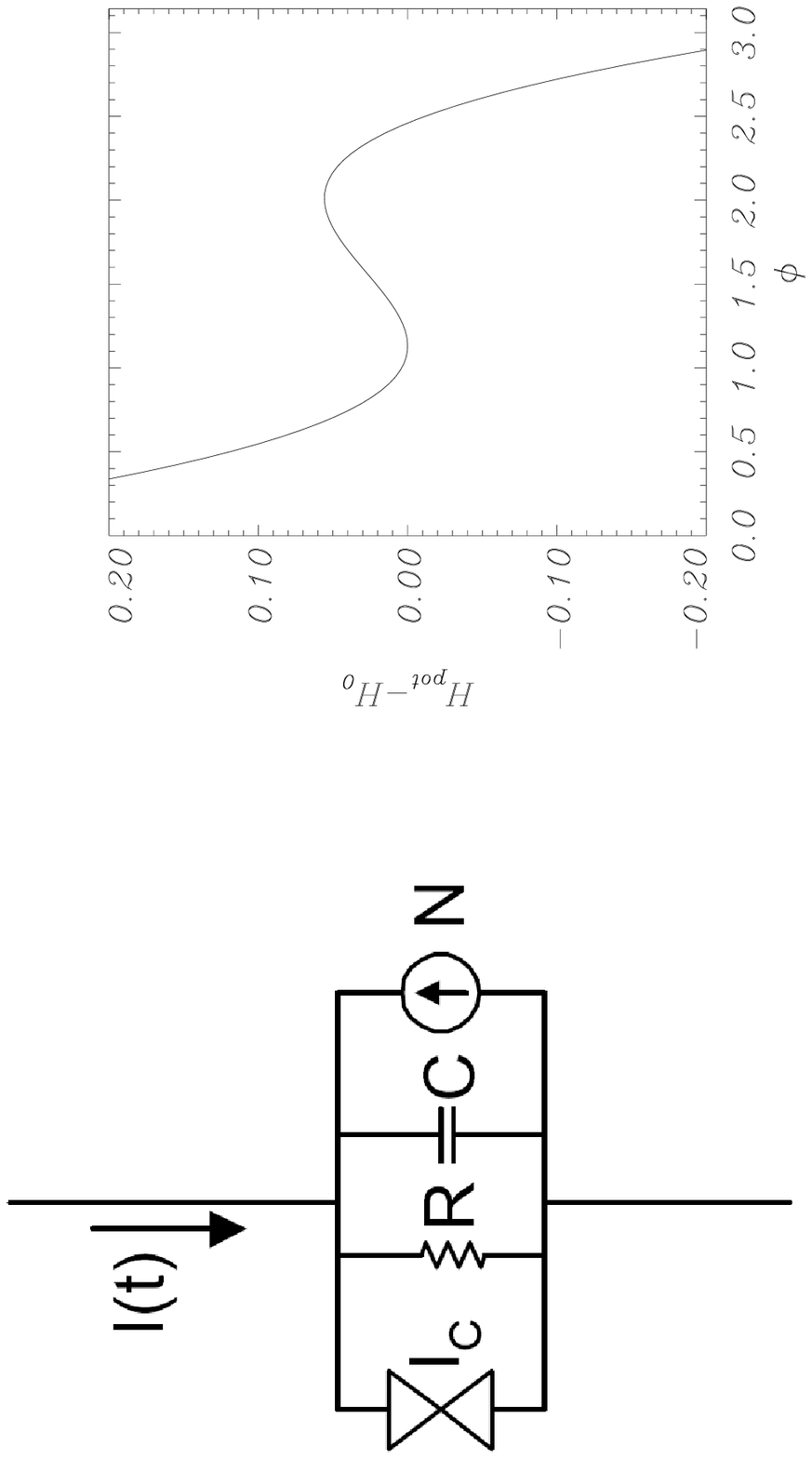}
    \end{picture}
\caption{Sketch of the RCSJ model where $I/I_c=\protect\eta+\protect%
\varepsilon_s(t)\sin(\protect\omega_st+\protect\theta)+\protect\varepsilon%
_p(t)$ and $N=n(t)I_c$. Also shown is the Josepshon potential for $\protect%
\eta=0.904706$.}
\label{fig:Fig1}
\end{figure}
\pagebreak 
\begin{figure}[h]
\setlength{\unitlength}{2.25 cm} 
\begin{picture}(28,8)(0,0)
        \includegraphics{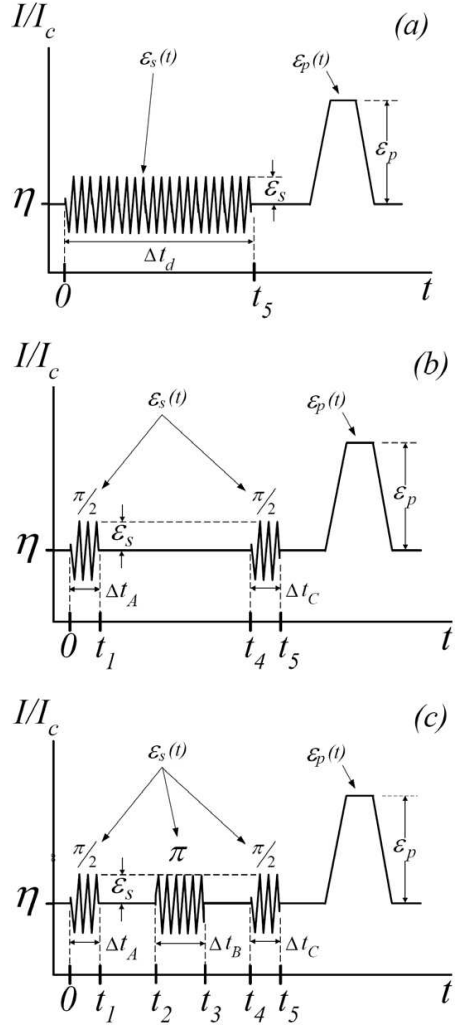}
    \end{picture}
\caption{Signalling diagrams for simulating (a) Rabi-type oscillations (b)
Ramsey-type fringes and (c) spin-echo-type oscillations. }
\label{fig:Fig2}
\end{figure}

\begin{figure}[tbp]
\setlength{\unitlength}{0.45 cm} 
\begin{picture}(30,22)(0,0)
        \includegraphics{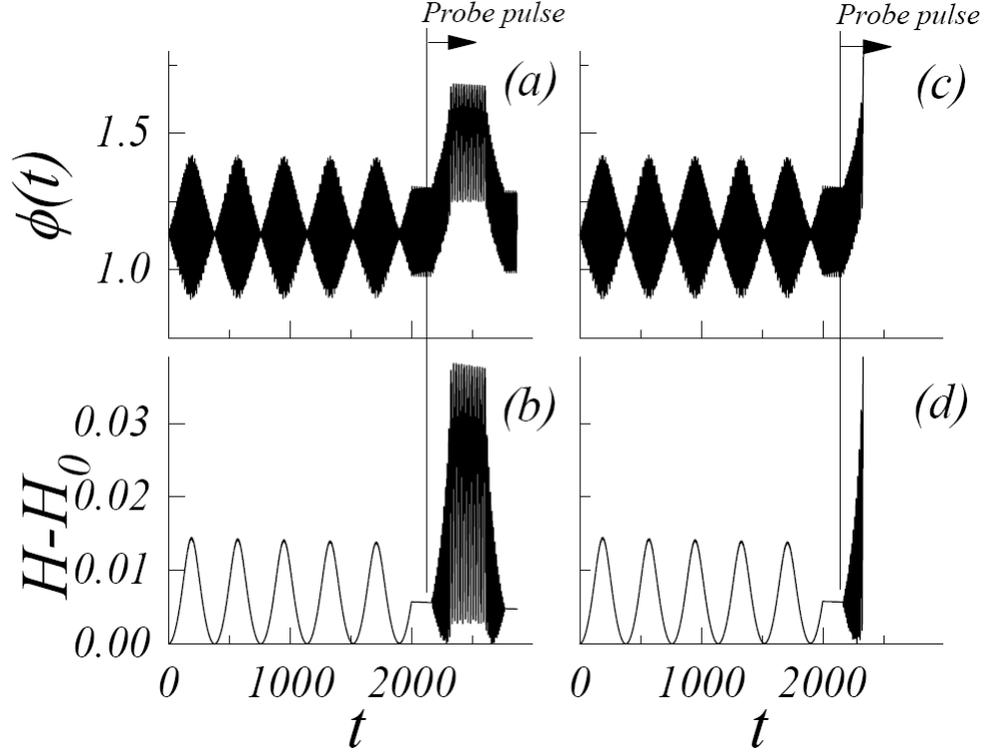}
    \end{picture}
\vspace{0.0 cm}
\caption{Direct simulation of the Rabi-type oscillation switching response.
Josephson junction response to the sequential application of two $\protect\pi%
/2$ microwave pulses followed by a probe field. Panels (a,b) show
phase-difference and energy for a non-switching sequence. Panels (c,d)
indicate a switching event. Switching and non-switching events trigger
according to a randomly determined phase. Parameters were $\protect\alpha%
=10^{-4}$, $\protect\eta=0.904706$, $\protect\varepsilon_s=2.17\times
10^{-3} $, $\protect\omega_s=\protect\omega_l=\sqrt[4]{1-\protect\eta^2}%
=0.652714$, $\protect\varepsilon_p=8.2\times 10^{-2}$, and $\Theta=0$. }
\label{fig:Fig3}
\end{figure}

\begin{figure}[tbp]
\setlength{\unitlength}{0.45 cm} 
\begin{picture}(35,28)(0,0)
        \includegraphics{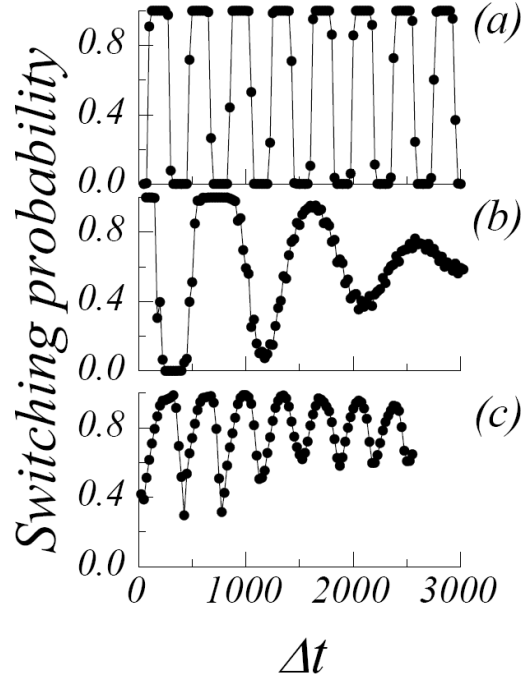}
    \end{picture}
\vspace{0.0 cm}
\caption{Switching probabilities for (a) Rabi-type oscillations, (b)
Ramsey-type fringe and (c) Spin-echo-type simulations. Each point represents
statistics of $\sim 2,500$ events at $\Theta=2.00 \times 10^{-4}$. The
horizontal axis in panel (a) represents the duration $\Delta t_d$ of the
microwave pulse. In panel (b) the horizontal axis indicates the $\protect\pi%
/2$-pulse separation ($t_4-t_1$). Panel (c) has the period between the end
of the first $\protect\pi/2$-pulse and the beginning of the $\protect\pi$%
-pulse -- i.e., the $\protect\pi$-pulse offset ($t_2-t_1$). In panels (a) \&
(b) the driving frequency is the linear resonance frequency $\protect\omega%
_s=\protect\omega_l=\sqrt[4]{1-\protect\eta^2}=0.652714$. In (c) the driving
frequency is $\protect\omega_s=0.965\protect\omega_l$. The magnitude of the
probe pulse in (a) \& (b) $\protect\varepsilon_p=8.2\times 10^{-2}$. The
probe pulse in panel (c) is $\protect\varepsilon_p=7.5\times 10^{-2}$. Panel
(c) reflects data from spin-echo-type simulations for $t_4-t_1=2750$. Common
to all panels are the characteristic damping $\protect\alpha=10^{-4}$ and
the microwave signal amplitude $varepsilon_s=2.17 \times 10^{-3}$. Lines are
added to aid the eye. }
\label{fig:Fig4}
\end{figure}

\begin{figure}[tbp]
\setlength{\unitlength}{0.45 cm} 
\begin{picture}(30,22)(0,0)
        \includegraphics{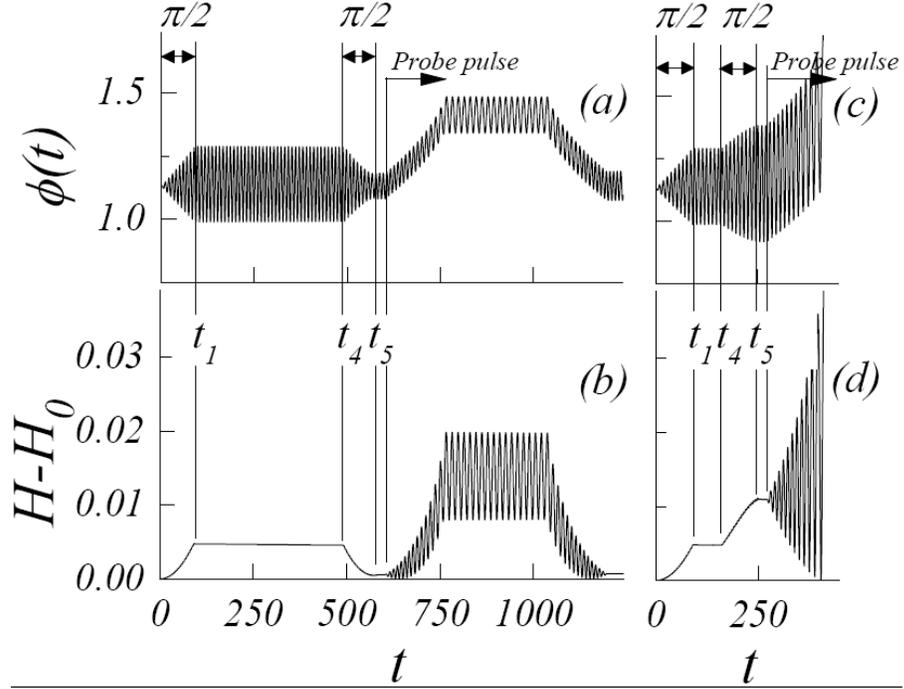}
    \end{picture}
\vspace{0.0 cm}
\caption{Direct simulation of the Ramsey-type fringe switching response.
Josephson junction response to the sequential application of two $\protect\pi%
/2$ microwave pulses followed by a probe field. Panels (a,b) show
phase-difference and energy for a non-switching sequence in the case of $%
t_4-t_1=400$. Panels (c,d) indicate a switching event for $t_4-t_1=70$.
Parameters were $\protect\alpha=10^{-4}$, $\protect\eta=0.904706$, $\protect%
\varepsilon_s=2.17\times 10^{-3}$, $\protect\omega_s=\protect\omega_l=\sqrt[4%
]{1-\protect\eta^2}=0.652714$, $\protect\varepsilon_p=8.2\times 10^{-2}$,
and $\Theta=0$.}
\label{fig:Fig5}
\end{figure}

\begin{figure}[tbp]
\setlength{\unitlength}{0.45 cm} 
\begin{picture}(35,28)(0,0)
        \includegraphics{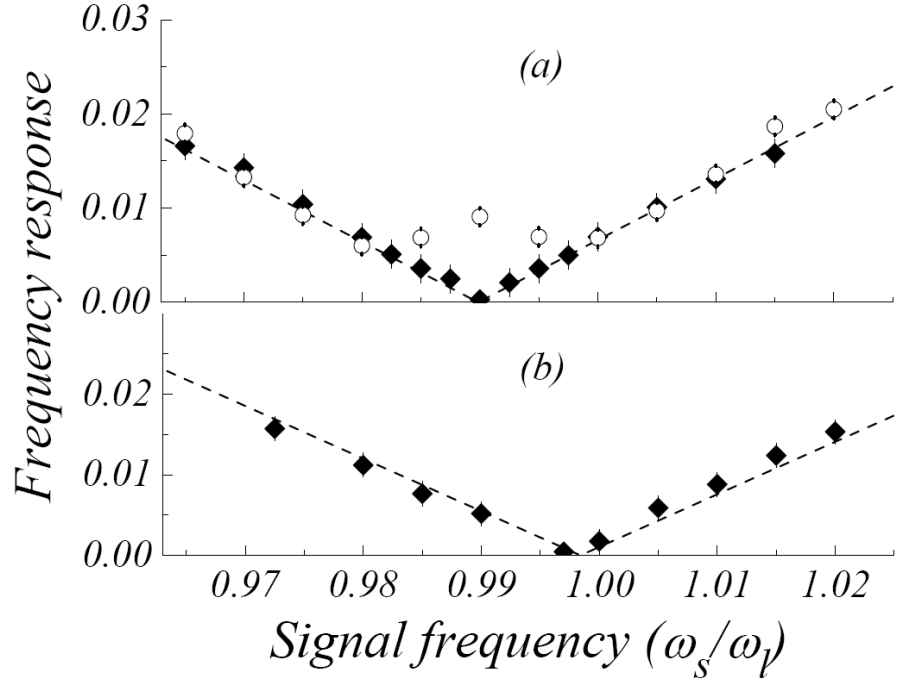}
    \end{picture}
\vspace{0 cm}
\caption{Frequency response as a function of applied microwave frequency for
two different dissipation parameters: (a) $\protect\alpha=10^{-4}$, and (b) $%
\protect\alpha=10^{-3}$. Filled diamonds depict Ramsey-type fringe
frequencies, $\Omega_F$, with parameters corresponding to Fig.~\protect\ref%
{fig:Fig4}. The open circles in (a) represent spin-echo-type frequencies
with parameters as in Fig.~\protect\ref{fig:Fig5}. }
\label{fig:Fig6}
\end{figure}

\begin{figure}[tbp]
\setlength{\unitlength}{0.45 cm} 
\begin{picture}(30,22)(0,0)
        \includegraphics{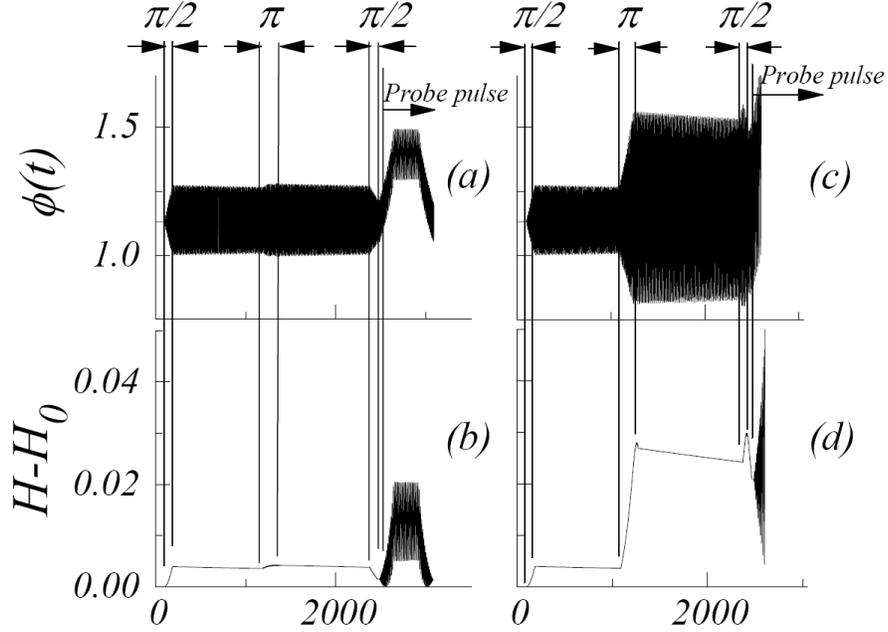}
    \end{picture}
\vspace{0.0 cm}
\caption{Direct simulation of the spin-echo-type oscillation switching
response. Josephson response to the application of two $\protect\pi/2$
microwave pulses, with an intervening $\protect\pi$-pulse, followed by a
probe field. The delay between the $\protect\pi/2$ pulses is $t_4-t_1=2182$.
The $\protect\pi$-pulse offset is for (a,b) (non-switching) $t_2-t_1=1000$.
For (c,d) (switching) $t_2-t_1=900$. Parameters were $\protect\alpha=10^{-4}$%
, $\protect\eta=0.904706$, $\protect\varepsilon_s=2.17\times 10^{-3}$, $%
\protect\varepsilon_p=7.5\times 10^{-2}$, and $\Theta=0$. }
\label{fig:Fig7}
\end{figure}

\end{document}